# Disability and Library Services: Global Research Trend

Swapan Kumar Patra

Department of Library and Information Science

Sidho-Kanho-Birsa University

Purulia, West Bengal, India

**Abstract**

The research on differently abled persons, and their use of library is getting global attention in recent years. The field has shown a modest, continuous but wide-scale growth. This research paper aimed at capturing the dynamics of the field using various bibliometrics and text mining tools. The bibliographic data of journal articles published in the field were collected from the Web of Science (WoS) database. The records were collected form the year 1991 to 2021 and analysed to observed the trends of literature growth, core journals, institutes from where most of the literature is being published, prominent keywords and so on. The results show that there is a significant growth of publications since the year 2000. The trends shows that the research in these areas is mostly emerging from developed countries. The developing countries should also pay more attention to do research in this area because differently abled peoples' need in developed countries may vary with respect to developed countries.

**Keywords:** Assistive technologies, Adaptive technologies, Library Services, Bibliometrics, Scientometrics, Social Network Analysis,

## Introduction

According to the United Nations (UN) Department of Economic and Social Welfare about 15 per cent of the world's population, or around 1 billion people in the world live with some form of disabilities[1]. World Health Organization (WHO) further observed that the number of people living with disability is increasing. Moreover, the growing ageing populations and various chronic health conditions are also responsible for the increasing disability in general population all over the globe[2]. Because of their inherent limitation of various kinds, this large population face various difficulties while getting the latest information in timely manners. Like all other, differently abled persons also require the easy access to information to meet their daily need, for their educational purpose and many other basic needs. Getting information in timely manner will empower a differently able person for getting employment and finally help them to be an independent individual.

It is universally recognised that differently-abled persons have the equal access rights of information. Governments all over the world has passed several acts or rules to ensure the rights of physically challenged persons. For example, Ministry of Social Justice and Empowerment, the Government of India has administered the 'Rights of Persons with Disabilities Act, 2016', 'The National Trust for the Welfare of Persons with Autism', 'Cerebral Palsy, Mental Retardation and Multiple Disabilities Act, 1999', 'Rehabilitation Council of India Act, 1992', 'the Persons with Disabilities Act 1995'. With all these acts, the government of India ensures equality, freedom, justice and dignity of all individuals and implicitly mandates an inclusive society for all, including the persons with disabilities. UN General Assembly adopted the convention to empower the Persons with Disabilities across the globe on their demands and their rights[3].

With this background this paper is an attempt to map the global research trends in this area using various bibliometric techniques. While doing so the paper has raised the following research questions: What are the most influential countries, institutions, sources doing research in this area. What types of documents published in the field of disability and library services? What are the hot research topics and the themes of research? What are the literature gaps and future research agendas?

The design of the paper is as follows: the next section will deal with the literature gap found through literature review. The section follows are the research objectives and methodology. Then the paper moves to the analysis part. The paper finishes with the concluding remarks.

## Literature review

The main theme of the search is the bibliometric analysis on "Assistive technology" (AT). The term encompasses the various devices, software, or equipment. These types of technologies are used to help people with various forms of disabilities. Using these technologies differently able person can work independently (Nandi, 2021). Moreover, assistive technologies are used to help persons with disabilities to make them more productive. AT can be either low technology or high technology depending on the use (Oyelude, 2017).

---

[1] https://www.un.org/development/desa/disabilities/resources/factsheet-on-persons-with-disabilities.html
[2] https://www.who.int/en/news-room/fact-sheets/detail/disability-and-health
[3] https://disabilityaffairs.gov.in/content/page/acts.php

Libraries around the world are making use of suitable assistive technologies to help their users with disabilities (Nandi, 2021). In the present information age, services offered by libraries globally depend on various types of AT to facilitate disabled patrons' information retrieval. Library patrons use AT for getting information for learning and teaching purpose. However, due to various factors sometimes the use of AT resulting in the underutilization of information services in libraries (Potnis & Mallary 2021).

However, there are a number of research articles have been published on the bibliometric analysis on Assistive Technologies (AT). Asghar, Cang & Yu (2017) observed that the United States of America (USA) was the leading contributor in the field of AT followed by UK. Although USA and UK are the two leading publishers in the field, however, publications from Germany have got more attention and relevance because of higher citation rate. The results showed that countries from the Europe were more active in collaborative research. The authors further concluded that the research outputs are highly correlated with the national policies and strategies adopted (dementia strategies) in those countries (Asghar, Cang & Yu 2017). Hence it can be concluded that research productivity is highly aligned with the national policy strategies.

Ryan et.al, (2004) conducted quantitative assessment of assistive technology from the MEDLINE database. The study identified core journals which are publishing significant numbers of papers related to AT and other related indicators. The study observed that the research papers in the areas of AT are scattered across a wide variety of journals. Moreover, some areas of AT have shown an increase in publications per year over time, while others have shown a more constant level of productivity (Ryan et.al., 2004).

A literature search was conducted to locate the scholarly literature in this area, using various indexing and abstracting databases. For example, literature from Web of Science (WoS), Scopus and Google Scholar were searched to find the bibliometric analysis of Disability and Library Services. It was observed that scholarly literature in this area is quite limited. Summers (1986) studied 2,270 scholarly journals on learning disabilities published between the years 1968 and 1983. The study traced the growth of research articles, core journals in the field and so on. This paper provided an important formal information source for researchers and practitioners (Summers 1986). However, the bibliometric study on disability and library service is quite limited.

With the brief literature review discussed above, the present study will try to fill the gap in the literature. The study deal with the following research objectives:

**Objectives**

The objectives of this research paper are:

- to study the trends in the growth of literature in this area
- to find the types of documents where the research items are being published
- to find the types of Web of Science subject categories and the trends in Keywords
- to identify the core journals in the field where the maximum number of papers being published.
- to find the most productive institutes and the country wise publication pattern.

## Methodology

A literature search was conducted on the Web of Science (WoS) database of Clarivate Analytics to retrieve the bibliographic records related to disability and the library services. The search terms were used to search and retrieve records where the AT and related terms are used. The search used the following search string: (((((((((AK= (Assistive technologies)) OR AK= (adaptive technologies)) OR AK= (Digital Talking Book)) OR AK= (disabilities information and communications technologies)) OR AK= (Library Service for the Blind)) OR AK= (Library Service Physically Handicapped)) OR AK= (braille book)) OR AK= (Library Service and Physically Handicapped)) OR AK= (Library Service and Handicapped)) OR AK= (Library Service AND Disable).

There are altogether 2,515 records retrieved for the period 1991-2021 using those search terms. The retrieved records were further analysed using various bibliometric / scientometrics indicators and the results are present using the following tables / graphs etc.

The social network maps were drawn using the Social Network Analysis tools. There are many open sources as well as commercial software available to facilitate and support the qualitative and quantitative analysis of social networks. Among the most popular software are Pajek, UCINET 6, NetDraw, Gephi etc. UCINET 6 is used to analyse multiple analytical tools. It is highly efficient for exploring and measuring social network structures. NetDraw, is a software tool nested in UCINET 6. Further Gephi is another social network software allows network visualization (Apostolato, 2013). UCINET, NetDraw and Gephi are used in this study to draw network map.

## Results

The following section will deal with the various results from the retrieved records. Different types of bibliometrics tools are used to get the indicators including the growth of literature, the document types of the retrieved literature, institutes form where most of the publications are coming, the core journals which publish about one third of publications, the productive institutes and their collaboration patterns, the prominent keywords of research and the country wise productivity and collaboration patterns.

**Growth of literature**

Using the search keywords mentioned above, there are altogether 2,515 records retrieved from the Web of Science (WoS) database during the period 1991-2021. There was only one publication in the year 1991. However, a visible growth of publication was observed from the year 2001 onwards. There were about 351 articles published in the year 2020 and 277 articles published (till October) in the year 2021. From the growth of literature curve (figure 1), it can be concluded that the publications trends in the area is linear with a visible growth in the recent years.

**Figure 1 Growth of literature**

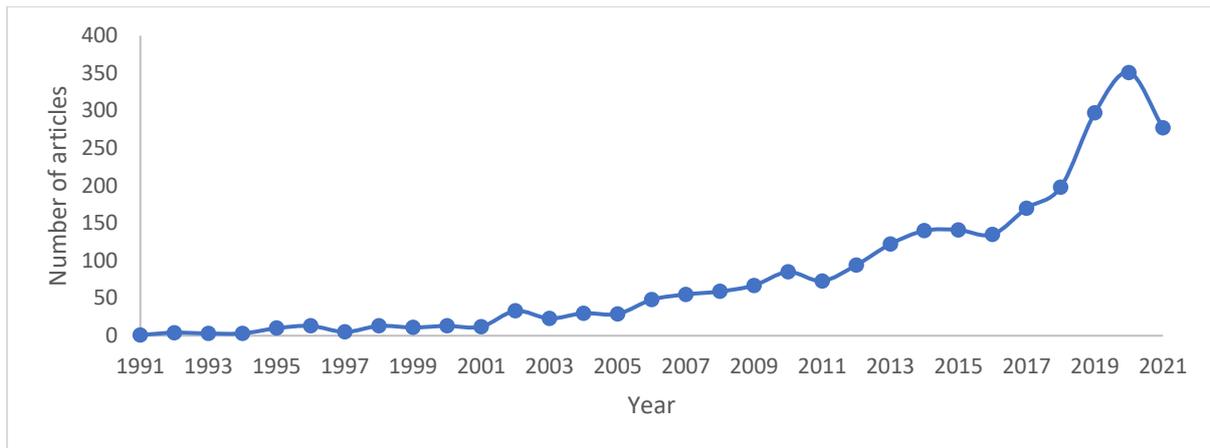

**Document types**

The document type analysis shows that maximum number of publications are the Articles (2,238; 88.99%), followed by Review Articles 204 (8.11%) Early Access 139 article (5.53%), Proceedings Papers 70 articles (2.78%), Meeting Abstracts 40 (1.59%) Editorial Materials 27 (1.07%), Book Reviews 2 articles (0.08%), Notes 2 articles (0.08%), Book Chapters 1 article (0.04%), Corrections 1 article (0.04%), Letters 1 article (0.04%). The graphical representations of the articles are shown in figure 2. So, from figure 2 it can be concluded that like any other filed of enquiry, journal publication is the most prominent form of publication in this area.

**Figure 2 Document types of publications**

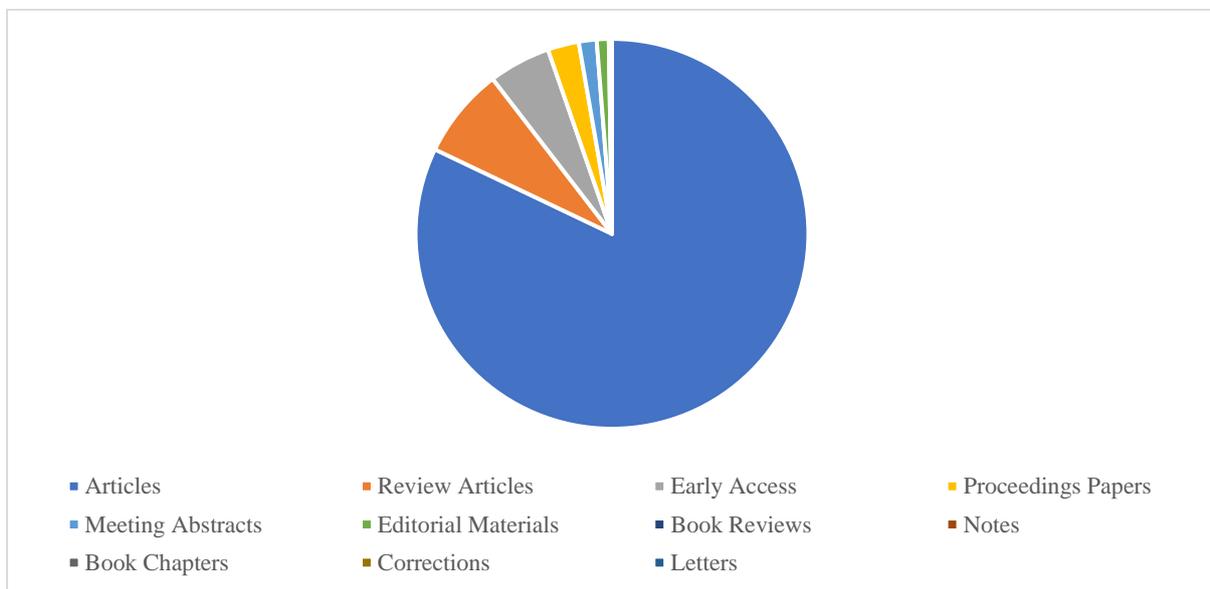

**Web of Science Categories**

Web of Science (WoS) have categorised the universe of subjects into 254 various subject categories. Every record in WoS collection contains the subject category of its source publication in the Web of Science Categories field. Table 1 shows the top subject categories of the retrieved documents. The top most subject category is the Rehabilitation (820 documents, 32.60%) followed by Engineering Electrical Electronic (323 documents, 12.84%), Computer Science Information Systems (190 documents, 7.56%), Engineering Biomedical (150 documents, 5.96%) and Telecommunications (131 documents 5.21%) and so on.

**Table1: Subject category of the retrieved documents**

| Sl. No | Web of Science Categories | Record Count | % of 2,515 |
|---|---|---|---|
| 1. | Rehabilitation | 820 | 32.60% |
| 2. | Engineering Electrical Electronic | 323 | 12.84% |
| 3. | Computer Science Information Systems | 190 | 7.56% |
| 4. | Engineering Biomedical | 150 | 5.96% |
| 5. | Telecommunications | 131 | 5.21% |
| 6. | Computer Science Cybernetics | 119 | 4.73% |
| 7. | Clinical Neurology | 112 | 4.45% |
| 8. | Computer Science Artificial Intelligence | 109 | 4.33% |
| 9. | Public Environmental Occupational Health | 107 | 4.25% |
| 10. | Education Special | 100 | 3.98% |
| 11. | Neurosciences | 99 | 3.94% |
| 12. | Ergonomics | 96 | 3.82% |
| 13. | Gerontology | 96 | 3.82% |
| 14. | Health Care Sciences Services | 95 | 3.78% |
| 15. | Instruments Instrumentation | 87 | 3.46% |
| 16. | Medical Informatics | 85 | 3.38% |
| 17. | Geriatrics Gerontology | 80 | 3.18% |
| 18. | Education Educational Research | 74 | 2.94% |
| 19. | Audiology Speech Language Pathology | 70 | 2.78% |
| 20. | Computer Science Interdisciplinary Applications | 70 | 2.78% |
| 21. | Information Science Library Science | 59 | 2.35% |
| 22. | Environmental Sciences | 54 | 2.15% |
| 23. | Engineering Multidisciplinary | 52 | 2.07% |
| 24. | Chemistry Analytical | 51 | 2.03% |

**Keyword Analysis**

Keyword analysis is an important research area in bibliometrics or scientometrics studies (Seale & Charteris-Black 2010). The derivation of new and valuable bibliometric indicators or approaches through keyword analysis is important for tracing the growth and the further development of the subject area (Wang & Chai 2018).

The author's keywords were downloaded and separated in the excel sheet for Keyword analysis. For keyword analysis, the social network analysis software Gephi was used. Gephi is an open interactive visualization and exploration platform, for all kinds of networks, dynamic and hierarchical graphs (Apostolato 2013). The network data was further exported to UCINET (Borgatti, Everett & Freeman, 1996) network file and NetDraw (Borgatti 2002) software was used to draw network map.

A total of 15,831 keywords are extracted from the total 2515 documents. The keywords formed a core cluster and periphery structure (figure 3). The whole network has 100 components and the largest component has the component size of 5592 element. "Assistive technologies" is the most prominent keyword and it form the largest component in the network. The second component consists of only 14 keywords and the most prominent keyword in that cluster is "multi-agent technology". Rest other clusters are very small and contains very minor keywords.

**Figure: 3 Author's Keyword clustering**

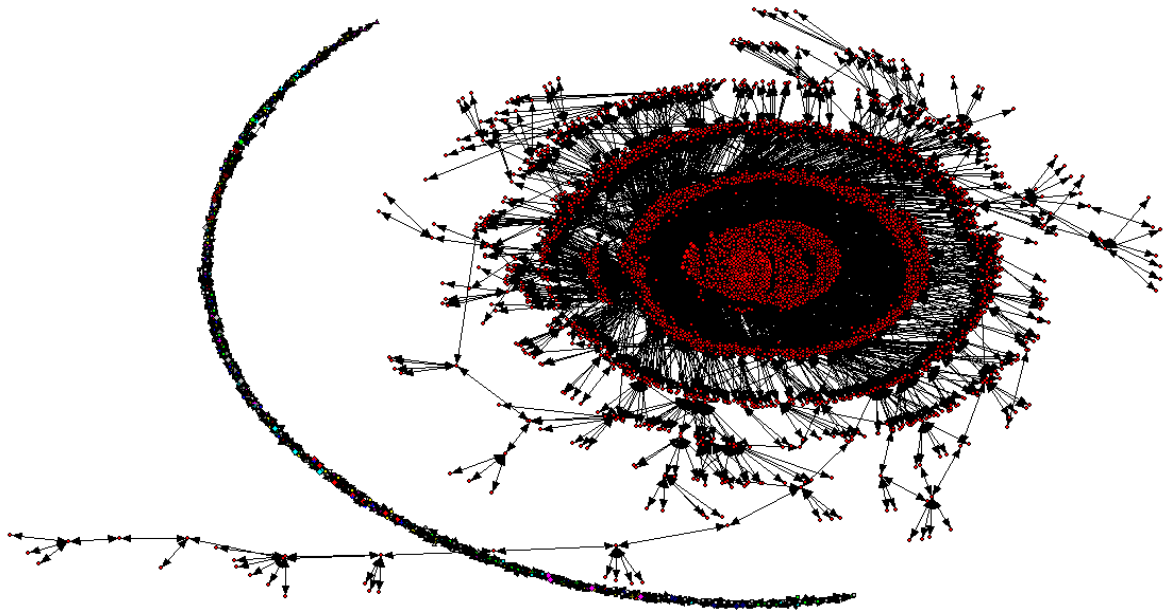

**Journals**

Journals are the most important and prominent outlet in scholarly publications. As seen from the document analysis about most of the publications in this research area is coming out as journal article. In this study, the total 2,591 retrieved records were published in 798 journals. A careful look at the publication trend (table 2) shows that the 21 journals altogether publish one third of total articles. Hence these journals can be considered as the "core" journals in this area.

**Table 2 Core journals**

| Sl. No. | Publication titles | Record count | Cumulative records | % of 2,515 |
|---|---|---|---|---|
| 1. | Disability and rehabilitation assistive technology | 146 | 146 | 5.805 |
| 2. | Assistive technology | 118 | 264 | 4.692 |
| 3. | IEEE access | 59 | 323 | 2.346 |
| 4. | IEEE transactions on neural systems and rehabilitation engineering | 50 | 373 | 1.988 |
| 5. | Disability and rehabilitation | 48 | 421 | 1.909 |
| 6. | Sensors | 48 | 469 | 1.909 |
| 7. | Journal of rehabilitation research and development | 44 | 513 | 1.75 |
| 8. | British journal of occupational therapy | 36 | 549 | 1.431 |
| 9. | Universal access in the information society | 35 | 584 | 1.392 |
| 10. | International journal of audiology | 32 | 616 | 1.272 |
| 11. | International journal of environmental research and public health | 31 | 647 | 1.233 |
| 12. | Journal of special education technology | 24 | 671 | 0.954 |
| 13. | Neurorehabilitation | 24 | 695 | 0.954 |
| 14. | Applied sciences basel | 18 | 713 | 0.716 |
| 15. | Archives of physical medicine and rehabilitation | 18 | 731 | 0.716 |

| 16. | Interacting with computers | 18 | 749 | 0.716 |
| 17. | Scandinavian journal of occupational therapy | 18 | 767 | 0.716 |
| 18. | Augmentative and alternative communication | 17 | 784 | 0.676 |
| 19. | BMC geriatrics | 16 | 800 | 0.636 |
| 20. | International journal of rehabilitation research | 16 | 816 | 0.636 |
| 21. | Journal of spinal cord medicine | 16 | 832 | 0.636 |

Bradford's law of scattering is used to identify the core journals in a given field (Patra & Chand, 2005; Patra & Chand 2006; Patra, Bhattacharya, & Verma 2006). This law is used to identify the core sets of journals, which publish the most contents of a given field. Bradford's law of scattering says that the scientific articles in a given field are generally concentrated in a "core" set of journals and they may be called as 'nucleus of periodicals more particularly devoted to the subject' (Vickery, 1948). Table 2 shows the Bradford core journals. These top 21 journals contain about 30 percent of the research output.

**Institute wise productivity**

Institute wise productivity patterns are shown in table 3. In the table top institutes are shown with more than 1 percent of publications. The institute wise productivity in the decreasing order are as follows; Pennsylvania Commonwealth System of Higher Education 92 articles (3.66%), University of Pittsburgh 61 articles (2.43%), US Department of Veterans Affairs 55articles (2.19%) and so on.

**Table 3 Institute wise productivity.**

| Sl. No | Name of the institute | Record count | % of 2,515 |
|---|---|---|---|
| 1. | Pennsylvania Commonwealth System of Higher Education | 92 | 3.66% |
| 2. | University of Pittsburgh | 61 | 2.43% |
| 3. | US Department of Veterans Affairs | 55 | 2.19% |
| 4. | University System of Georgia | 54 | 2.15% |
| 5. | Veterans Health Administration | 54 | 2.15% |
| 6. | University of London | 52 | 2.07% |
| 7. | University of Montreal | 50 | 1.99% |
| 8. | State University System of Florida | 48 | 1.91% |
| 9. | University of Toronto | 48 | 1.91% |
| 10. | University of Texas system | 40 | 1.59% |
| 11. | Universita degli studi di bari aldo moro | 33 | 1.31% |
| 12. | University of California System | 33 | 1.31% |
| 13. | Georgia Institute of Technology | 30 | 1.19% |
| 14. | University of British Columbia | 29 | 1.15% |
| 15. | Duke University | 28 | 1.11% |
| 16. | University of Michigan | 28 | 1.11% |
| 17. | University of Texas Austin | 27 | 1.07% |
| 18. | Pennsylvania State University | 26 | 1.03% |

It is observed from the table 3 that most of the journal articles are coming from the universities / institutes of US. There are no university research institute or other institutes have good number of publications from the developing countries.

**Institutional collaboration patterns**

UCINET 6, a Windows software package used for the analysis of social network data. The software has been developed by Steve Borgatti, Martin Everett and Lin Freeman (2002). This package provides the tools to analyze 1-mode or 2-mode data. This software was used to find various actor level centrality measure (Apostolato 2013). NetDraw is a program developed by Borgatti (2002) for visualizing 1-mode and 2-mode social network data.

A social network can be defined as a network of various social entities (actors) and the linkages (relations) among them (Figure 4). This study studied the institute wise collaboration pattern through the co authored articles by the authors from different institutes. It is assumed that if the authors co-authored article means they have some form of collaboration among them. The one mode network graph is drawn using the Social Network Analysis tools.

**Figure 4 Institutional collaboration patterns**

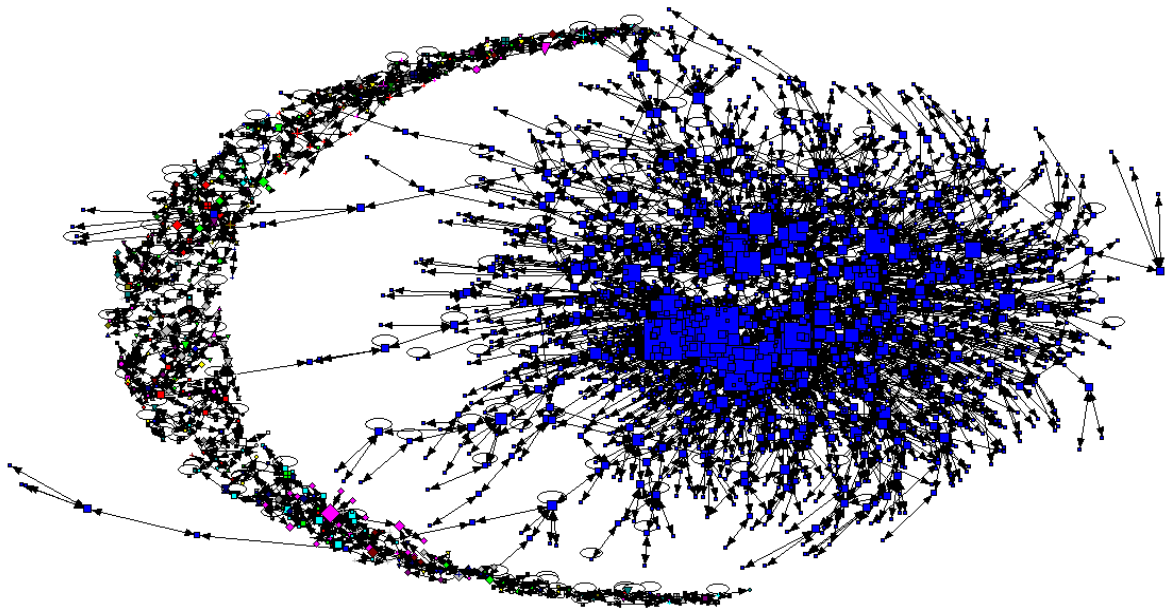

Here, the individual actor level collaboration patterns are measures using various centrality measures for example degree centrality, betweenness centrality, closeness centrality and eigenvector centrality. The term *degree centrality* refers to a measure of the prominence or importance of an individual actor within a social network. It is measured by the total amount of direct links with the other nodes (Zhang & Luo 2017). In institute level collaboration pattern this study will only consider the degree centrality measure to show the strength of collaboration of various institutes (Table 4).

**Table 4 Degree centrality measure of collaborating institutes**

| Sl No | Name of the Institute | Degree |
|---|---|---|
| 1. | University of Pittsburgh | 40 |
| 2. | University of Montreal | 37 |
| 3. | University of Toronto | 31 |
| 4. | University of British Columbia | 30 |
| 5. | Maynooth University | 22 |

| 6. | University of Bari | 21 |
|---|---|---|
| 7. | University Florida | 20 |
| 8. | Georgia Institute of Technology | 20 |
| 9. | UCL | 20 |
| 10. | University of Manchester | 19 |
| 11. | Graz University of Technology | 19 |
| 12. | University of Sydney | 18 |
| 13. | Duke University | 17 |
| 14. | University of Michigan | 17 |
| 15. | University of Alberta | 16 |
| 16. | University of Western Ontario | 15 |
| 17. | Pennsylvania State University | 15 |

**Country wise publication pattern.**

The country wise publication profile shows that USA is the largest producer of research articles (837 documents ,33 percent) in this area followed by UK and Canada. Among the developing countries China is in the 4th position with 150 articles (about 6 percent). Brazil is in 11th position and India is in 13th position.

**Table 5 Country wise publication pattern.**

| Sl. No | Field: | Record Count | % of 2,515 |
|---|---|---|---|
| 1. | USA | 837 | 33.28% |
| 2. | England | 296 | 11.77% |
| 3. | Canada | 210 | 8.35% |
| 4. | Peoples R China | 150 | 5.96% |
| 5. | Italy | 145 | 5.77% |
| 6. | Australia | 133 | 5.29% |
| 7. | Germany | 121 | 4.81% |
| 8. | Spain | 111 | 4.41% |
| 9. | Sweden | 100 | 3.98% |
| 10. | Netherlands | 92 | 3.66% |
| 11. | Brazil | 80 | 3.18% |
| 12. | Ireland | 67 | 2.66% |
| 13. | India | 64 | 2.55% |
| 14. | Japan | 57 | 2.27% |
| 15. | Scotland | 55 | 2.19% |

**Concluding Remarks**

This study conducted a bibliometric analysis of Disability and Library Services. The study mainly focussed on the various disability services and their uses in the library. The use of Assistive technology growing at the global level. Libraries all over the world are increasingly using this technology in their daily operations at different levels. The study used various bibliometric and social network tools to analyse the trends in research. However, it is observed that the research on these technologies are still a developed countries phenomenon and very little contribution from the developing part of the globe. The further study recommends more in-depth analysis of this trend particularly in developing country's context. The developing countries should also pay more attention to do research in this area because differently abled peoples' need in developed countries may vary with respect to developed countries.